\documentclass[11pt]{article}
\usepackage{moriond,epsfig}

\pdfoutput=1

 \usepackage{graphicx}                   

\def\Journal#1#2#3#4{{#1} {\bf #2}, #3 (#4)}

\def\JHEP{\em J. High Energy Phys.}
\def\JPG{\em J. Phys. G Nucl. Part. Phys.}

\def\NIMA{{\em Nucl. Instrum. Methods Phys. Res.} A}

\def\PRD{{\em Phys. Rev.} D}

\def\be{\begin{equation}}
\def\ee{\end{equation}}
\def\bea{\begin{eqnarray}}
\def\eea{\end{eqnarray}}

\begin{document}
\vspace*{4cm}
\title{BEYOND THE SM SCALAR BOSON SEARCHES AT THE TEVATRON}

\author{Elem\'er NAGY }

\address{Centre de Physique des Particules de Marseille, 163 avenue de Luminy, Case 902,\\
F-13288  Marseille CEDEX 09, France}

\maketitle\abstracts{Recent results from the Tevatron are reported 
on Higgs boson searches in models beyond the standard model (SM). 
The models include fermiophobic Higgs bosons, 
the extension of the SM to a fourth 
generation of fermions, supersymmetric scenarios
and heavy Higgs boson cascade decays.
}

\section{Introduction}\label{sec:intro}

Recent results on Higgs boson searches beyond
the standard model (SM) are presented 
on behalf of the CDF and D0 collaborations.   
The data were collected at the Tevatron, 
a proton-antiproton collider with
1.96 GeV center of mass energy, 
using two general purpose detectors.
Both detectors had similar structure with 
different particular advantages.
While the CDF detector~\cite{CDFdet} had a larger volume for tracking
of the charged particles, the D0 detector~\cite{D0det} had a hermetic liquid
argon calorimeter and a muon detector with larger coverage
inside an iron toroidal magnet. The Tevatron operation
stopped on September 2011 after 10 years of running, delivering
about 12~fb$^{-1}$ integrated luminosity per experiment providing
about 10~fb$^{-1}$ of analyzable data for each of the collaborations.

\section{Higgs boson searches in the extension of the SM 
to a fourth generation of fermions}\label{sec:4gen}

A fourth generation of fermions in the SM is an interesting
possibility, since it is not ruled out by precision electroweak
data and it opens up new sources of CP violation. Moreover,
in this model (SM4) the production cross section 
of the Higgs boson is enhanced
by a factor of about 9 due to the additional heavy quarks in
the fermion loop of the gluon-gluon fusion (ggH) which becomes
an overwhelmingly dominant production process.
The CDF and D0 experiments used in this study the 
event selection designed for the
SM Higgs boson searches in the $WW$ and $ZZ$ final states~\cite{SMH} extending the
Higgs boson mass range up to 300 GeV. In doing so, they reoptimized
the separation of the signal from the background since,
contrary to the SM Higgs boson search, here 
both the vector boson fusion (VBF) and the associate production
of the Higgs particle with a vector boson (VH) are ignored. 

\begin{figure}[h]
\begin{center}
  \mbox{\includegraphics[width=10cm]{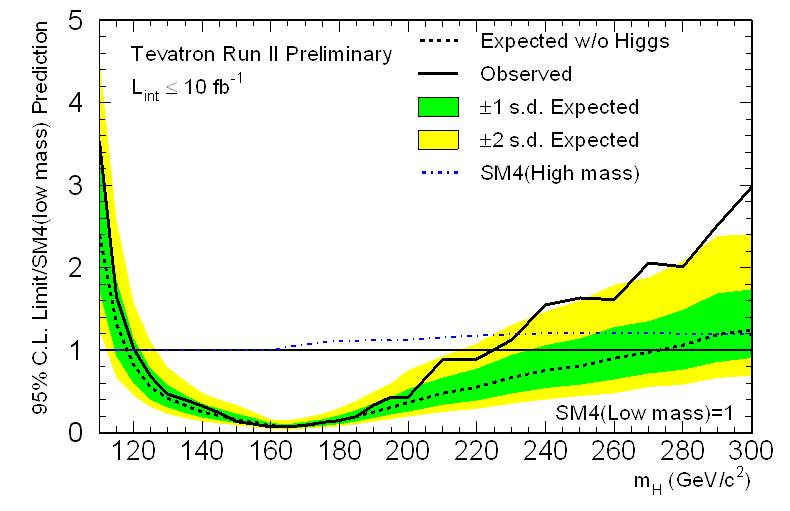}}
\caption{Observed (solid line) and expected (dotted line) 95\% CL 
cross section times $BR$ upper limits
of the Higgs boson as a function of its mass, 
in the SM4 model. 
The limit is a combination of the CDF and D0 measurements 
on the full dataset,
expressed in units of the theoretical cross section of the
low mass scenario. Also shown is in the same unit  
the theoretical cross section of the high mass scenario (dash-dotted line).
The green and yellow shaded area indicate the 68\% and 95\% probability
regions where the limits can fluctuate, in the absence
of signal.
}
\label{fg:4GLimit}
\end{center}
\end{figure}

Since no excess was observed above the background
expectation, a limit was set on the cross section of the Higgs particle  
produced in ggH and decaying into a $WW$ pair in the SM4 model,
assuming that the ratio of the branching fractions 
$BR(H\longrightarrow WW)/BR(H\longrightarrow ZZ)$ is the same
as in the SM. Two scenarios have been considered: in the low mass scenario
the fourth generation charged and neutral lepton masses are close 
to their experimentally determined lower bounds: $m_{l4}=100$ GeV
and $m_{\nu 4}=80$ GeV, whereas in the high mass
scenario they are both equal to 1 TeV.
In both cases the fourth generation quark masses are set to 
$m_{u4}=450$ GeV and $m_{d4}=400$ GeV.
Figure~\ref{fg:4GLimit} shows the combined observed and expected
cross section times $BR$ upper
limits at 95\% CL, expressed in units of the theoretical cross section of the
low mass scenario. From there the following
mass ranges can be excluded for the Higgs boson in the SM4 model:
120--224 GeV (observed), 118--274 GeV (expected) and 
120--232 GeV (observed), 118--291 GeV (expected) in the
low and high mass scenarios, respectively. 

\section{Fermiophobic Higgs boson searches}\label{sec:FHM}

\begin{figure}[h]
\begin{center}
  \mbox{\includegraphics[width=10cm]{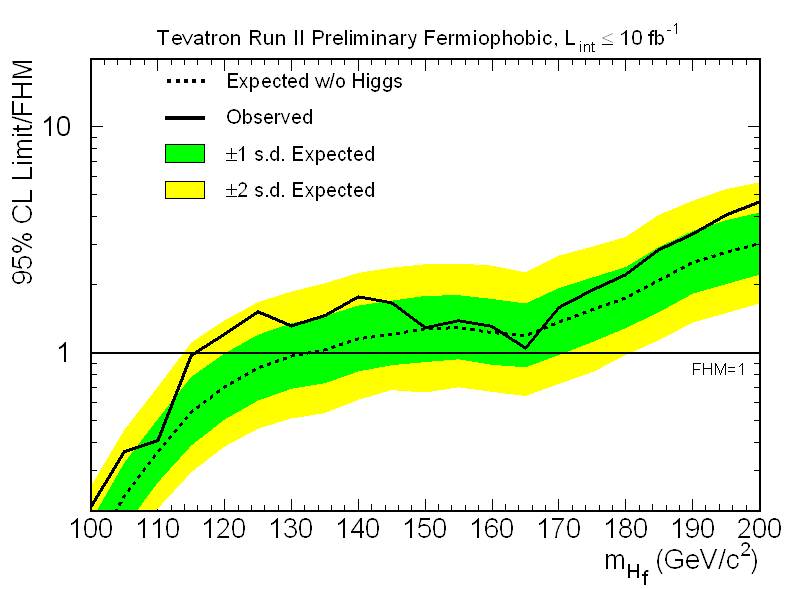}}
\caption{Observed (solid line) and expected (dotted line) 95\% CL 
cross section times $BR$ upper limits of the fermiophobic Higgs boson 
as a function of its mass. 
The limit is a combination of the CDF and D0 measurements
on the full dataset, 
expressed in units of the FHM theoretical prediction. 
The green and yellow shaded area indicate the 68\% and 95\% probability
regions where the limits can fluctuate, in the absence
of signal.  
}
\label{fg:FFHLimit}
\end{center}
\end{figure}

In the Fermiophobic Higgs Model (FHM), 
one assumes that the coupling of the Higgs boson to
fermions vanishes and all other couplings remain the same as
in the SM. This scenario can arise in models with an extended Higgs 
sector like a two Higgs Doublet Model (2HDM)
with parameters that make the lightest Higgs boson fermiophobic~\cite{FFH}.
A fermiophobic Higgs boson is dominantly produced via
VH and VBH. Moreover, its decay into two photons is largely enhanced,
such that this decay mode provides the best search sensitivity
for Higgs boson masses below 120 GeV. The CDF and D0 collaborations
therefore reinterpreted
the SM Higgs boson searches in the $\gamma\gamma$ and $WW$
final states~\cite{SMH}. They reoptimized the signal separation
from the background to account for the absence of the ggH production process.
Figure~\ref{fg:FFHLimit} shows the combined 
observed and expected cross section times $BR$ upper limits at 95\% CL,
in units of the FHM theoretical prediction. 
From there one can exclude  100--116 GeV (observed) and 
100--132 GeV (expected) mass ranges for a fermiophobic Higgs boson. 

\section{Search for $\Phi\longrightarrow b\bar b$ in MSSM}\label{sec:3b}

In the minimal supersymmetric extension of the SM (MSSM)
there are two complex Higgs doublet fields from which
five Higgs bosons arise after the electroweak symmetry breaking:
three neutrals ($h$, $H$, $A$), commonly denoted as $\Phi$
and two charged one ($H^{\pm}$). At tree-level, the model is
fixed by two parameters: $\tan\beta$, the ratio of the vacuum
expectation value of the two Higgs doublet fields and $M_A$,
the mass of the CP-odd neutral Higgs boson. The other model
parameters enter through radiative corrections. The mass
of the lightest neutral Higgs boson, $m_h$ has an upper bound.
For $\tan\beta>1$, the coupling of the $\Phi$ 
to down-type fermions becomes large
and therefore it decays with about 90\% branching fraction to
a $b\bar b$ pair. Moreover, the associate production of the
$\Phi$ with $b$ quarks is enhanced by a large factor ($\sim 2 \tan^2\beta$)
with respect to the SM Higgs production. CDF and D0 therefore searched for 
the $\Phi$ boson as a resonant peak in the di-jet invariant mass distribution
of events with 3 or 4 $b$-tagged jets.  

Using 2.6 fb$^{-1}$ of data, CDF selected $\sim$11 500 events with 3 $b$-tagged jets.
D0 analysed 5.2 fb$^{-1}$ of data resulting in $\sim$15 000 and $\sim$11 000 events
with 3 and 4 $b$-tagged jets, respectively.
Both experiments used PYTHIA~\cite{PYTHIA} to generate signal events 
subsequently weighted by MCFM~\cite{MCFM}, 
and estimated the multijet background from data. 
CDF enhanced the $b$-tagging algorithm by an additional flavour separator 
based on the invariant mass of the charged particles issued from the secondary vertex.
D0 used a likelihood ratio discriminant to augment
the separation of the signal from the background.
\begin{figure}[t]
\begin{center}
  \mbox{\includegraphics[width=7.5cm]{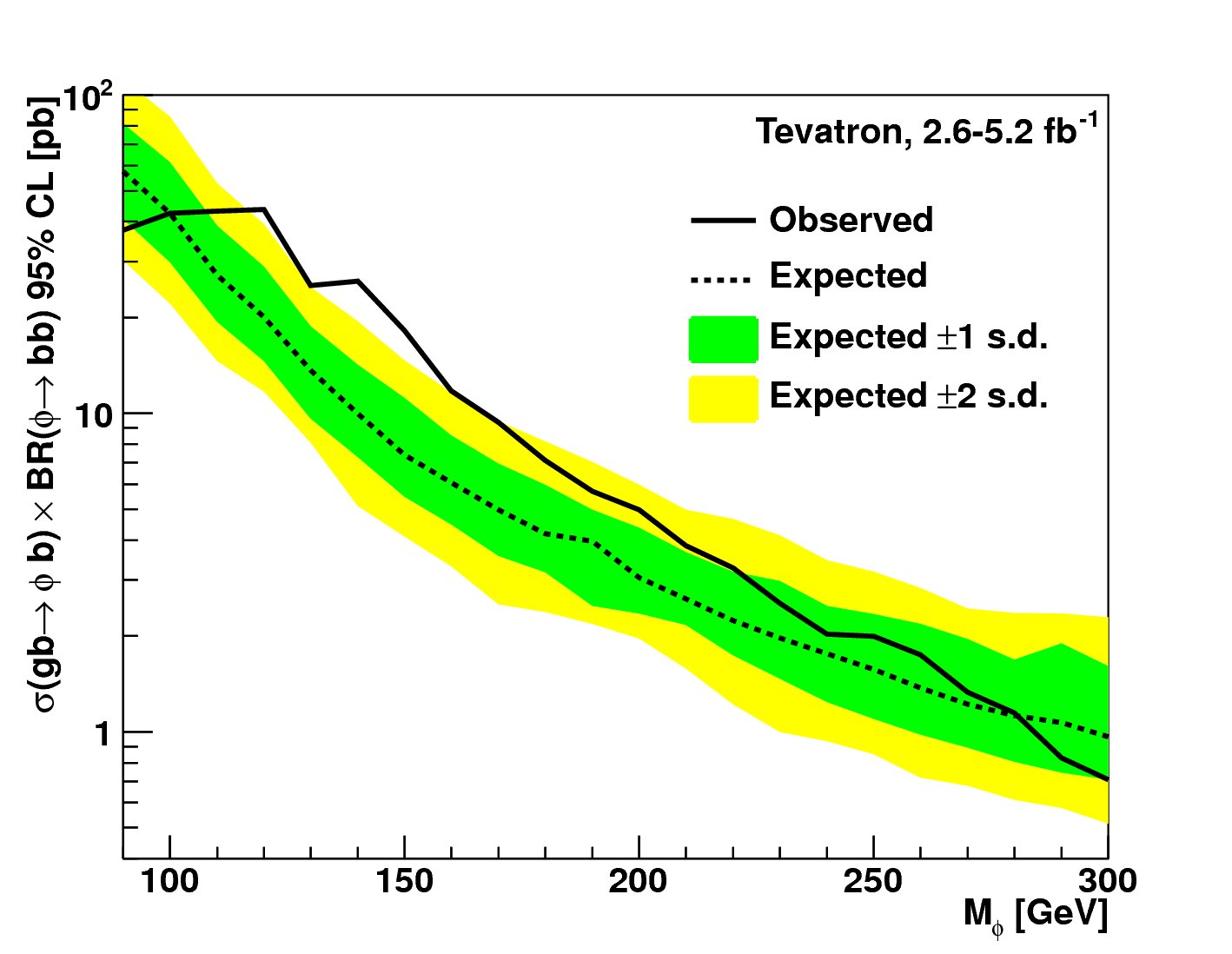}}
\caption{Observed (solid line) and expected (dotted line) 95\% CL 
cross section times $BR$ upper limits of the $\Phi$ boson 
as a function of its mass, produced in association with
$b$ quarks and decaying in $b\bar b$ quark pairs. 
The limit is a combination of the CDF and D0 measurements. 
The green and yellow shaded area indicate the 68\% and 95\% probability
regions where the limits can fluctuate, in the absence
of signal.    
}
\label{fg:tev-3b-xsec}
\end{center}
\end{figure}
\begin{figure}[h]
\begin{center}
  \mbox{\includegraphics[width=7cm]{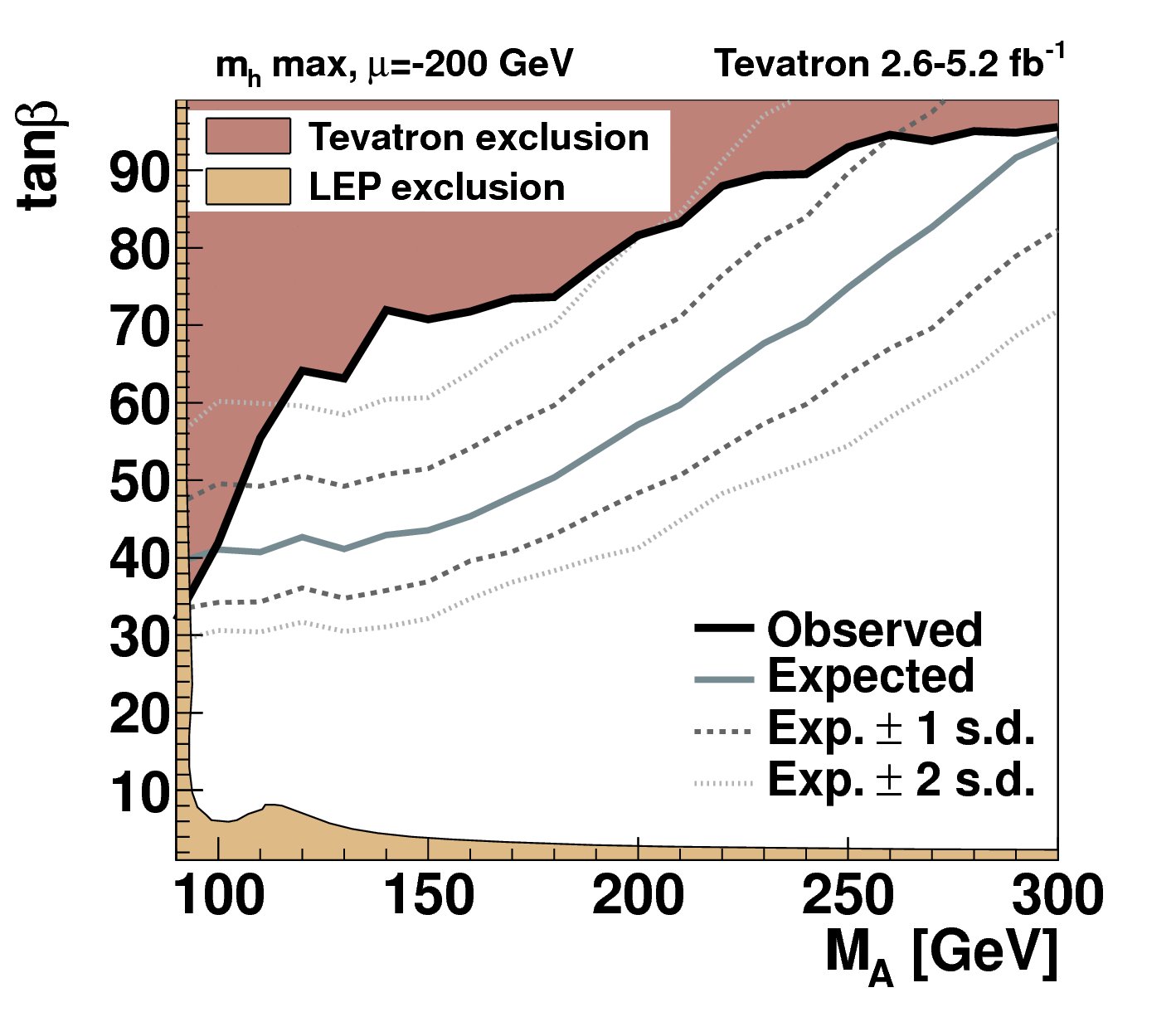}}
\caption{Excluded region in the ($M_A,\tan\beta$) plane (dark shaded area)
obtained in the $m_h^{max}$ scenario. Also shown are the excluded
region by the LEP experiments (light shaded area) as well as the
median expected upper limits of $\tan\beta$ {\em vs} $M_A$ (solid line).
The dashed and dotted lines around the solid line
enclose the 68\% and 95\% probability
regions where the limits can fluctuate, in the absence
of signal.    
}
\label{fg:tev-3b-mssm}
\end{center}
\end{figure}
Since no significant resonant peak was found by either experiment,
a combined 95\% CL upper limit of the $\Phi$ production cross section
times $BR$ was determined (Figure~\ref{fg:tev-3b-xsec}).
No radiative corrections were taken into account and the width of the $\Phi$
boson was neglected. The local excesses seen in the observed limit at 120 GeV and
140 GeV correspond to 2 standard deviations after applying trial factors
which take into account the number of mass regions investigated. 
In addition, excluded regions in the $\tan\beta$ {\em vs} $M_A$
were determined for different MSSM model parameters
applying radiative corrections and taking into account the width
of the $\Phi$ boson. Figure~\ref{fg:tev-3b-mssm}
shows the so-called $m_h^{max}$ scenario, where
the parameters were chosen to maximize the upper bound of $m_h$.
These results were published in 2012~\cite{TeV3b} and represented
the best limits and excluded regions until 
the CMS collaboration has superseded it recently~\cite{CMS3b}. 


\section{Search for heavy Higgs boson cascade decays}\label{sec:Hcasc}

\begin{figure}[h]
\begin{center}
  \mbox{\includegraphics[width=12cm]{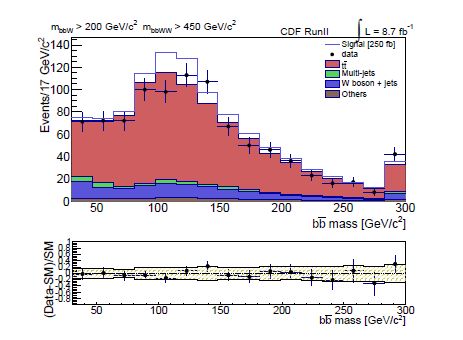}}
\caption{Invariant mass distribution of reconstructed $b$-tagged 
jet pairs for observed data and expected backgrounds. 
A signal hypothesis is shown, assuming a total cross 
section of 250 fb, 500 GeV and 300 GeV for the masses
of $H^0$ and $H^{\pm}$, respectively.
The lower panel shows the relative difference between 
the observed and expected distributions with the 
combined statistical and systematic uncertainties 
of the expected background.
}
\label{fg:H0CascadeMbb}
\end{center}
\end{figure}

The CDF collaboration searched for a hypothetical heavy 
neutral Higgs boson ($H^0$) 
which would first decay to a medium heavy charged Higgs boson 
($H^{\pm}$) and a $W$ boson. The $H^{\pm}$ then would decay into 
a light neutral Higgs
boson ($h^0$) of mass of 126 GeV and a second $W$ boson. Finally,
the $h^0$ would turn into a $b\bar b$ quark pair. This search is
motivated by a possible existence of strongly coupled
electroweak symmetry breaking sector in extended Higgs sectors,
like 2HDM~\cite{HCasc}.
Since the final state is similar to a $t\bar t$ pair production,
the same event selection was used as in the $t\bar t$
lepton+jets analyses. The signal was generated with
MADGRAPH~\cite{MADG} interfaced with PYTHIA.
The dominant backgrounds ($t\bar t$ and $W+$jets) were simulated
with ALPGEN~\cite{ALPG} interfaced with PYTHIA and the multijet
background was estimated from data. The reconstruction of the
decay chain started with the reconstruction of the $W$ bosons
from the untagged jet pairs, the signal is then searched in the
invariant mass distribution of the
$b$-tagged jet pairs (Figure~\ref{fg:H0CascadeMbb}).
As no significant excess of the signal was seen in this distribution,
upper limits for the production cross section times $BR$
were determined as a function of the $H^0$ and $H^{\pm}$
masses. These limits, however, exceed the corresponding
theoretical values, therefore no exclusion region could be derived
for the masses of the heavy neutral and charged Higgs bosons.
More details can be found in the public document~\cite{HCasc}.

\section{Summary}\label{sec:Sum}

Searches were presented for Higgs bosons beyond the SM, 
carried out by the CDF and D0 collaborations. No such signals
have been observed. Mass ranges have been excluded for
Higgs bosons assuming a fourth generation of fermions and for
fermiophobic Higgs bosons, using the full available
dataset collected at the Tevatron by the two experiments~\cite{TeVCombo}. 
Upper limits have been derived and domains in the MSSM planes
have been excluded for associate production of Higgs
bosons with $b$ quarks and decaying into $b \bar b$
quark pairs. Finally, upper limits for the production
and cascade decay of a heavy Higgs boson were derived
in a particular model.  

More details can be obtained from the CDF and D0 public web pages: 
\newline
http://www-cdf.fnal.gov/physics/new/hdg/Results.html
\newline
http://www-d0.fnal.gov/Run2Physics/WWW/results/higgs.htm

\section*{Acknowledgments}
In preparing this talk advices from the Physics Coordinators and Higgs
group conveners of the CDF and D0 collaborations were greatly appreciated.
The wonderful environment and the remarkable work of the Organizers
made this meeting a memorable event.

\end{document}